\documentclass{article}
\usepackage{amsmath,amssymb}
\usepackage{epsfig,multirow}
\usepackage{xcolor}
\usepackage{bigstrut}
\usepackage{array}
\usepackage{hyperref}
\usepackage{cite}

\usepackage[a4paper,top=3cm,bottom=2cm,left=3cm,right=3cm,marginparwidth=1.75cm]{geometry}
\title{Impact of the EPOS.LHC-R hadronic interaction model on the Centaurus~A ultrahigh-energy cosmic-ray scenario}

\author{Silvia Mollerach and Esteban Roulet\\
Centro At\'omico Bariloche, Comisi\'on Nacional de Energ\'\i a At\'omica\\
Consejo Nacional de Investigaciones Cient\'\i ficas y T\'ecnicas (CONICET)\\
Av. Bustillo 9500, R8402AGP, Bariloche, Argentina}
\date{}

\begin{document}
\maketitle
\begin{abstract}
    We discuss the impact of the recent EPOS.LHC-R hadronic interaction model on the scenario in which most of the cosmic rays with energies above 5~EeV originate in the nearby Centaurus~A radiogalaxy. The heavier composition inferred from this hadronic model has important implications for the interpretation of the spectral features and for the anisotropies. In particular, the amount of H and He present above the ankle is now very suppressed. The elements of the CNO group from the source play a  predominant role in the instep region just above the ankle, while the elements of the Si and Fe groups contribute significantly in the suppression region of the energy spectrum above 50~EeV. The lack of He from the source above the ankle also leads to a smaller expected anisotropy around the Cen~A direction at energies of 10 to 20~EeV. The expected dipolar anisotropy in different energy bins above 4~EeV is well consistent with the measurements for appropriate values of the extragalactic turbulent magnetic field and source lifetime. The overall fit to the different observables improves considerably with the new hadronic model.
   A new method is introduced to extract the information on the cosmic ray masses from that of the depth of the maximum development of the air showers, which significantly improves the existing method.

\end{abstract}
\section{Introduction}
One of the main open problems associated with the ultrahigh-energy cosmic rays (UHECRs) is the identification of their sources. Among the indications we have to shed light on this issue is the observed dipolar anisotropy in their arrival directions, which has an amplitude of about 6\% at 10~EeV that grows with increasing energy \cite{aa17,aa18}, as well as hints of a more localized excess appearing above 40~EeV in a window of about 30$^\circ$ radius close to the direction towards the nearby galaxy Centaurus~A (Cen~A, which is the radio counterpart of NGC~5128) \cite{ab22}. This is the closest radiogalaxy to us, at just about 4~Mpc distance, and it displays powerful jets and outflows which are due to the strong nuclear activity that likely started after the collision of two galaxies that happened less than 2~Gyr ago. This source has been proposed for a long time as a potential contributor to the highest energy cosmic rays, see e.g. \cite{gi80, be90,ro96, bl99,fa00,is01, ma18}. In ref.~\cite{mo24} we put forward a scenario in which most of the cosmic rays (CRs) observed above  5~EeV (the ankle energy) come from this source. This scenario considers a composition in agreement with that inferred from the observations of the depth of maximum development of the air showers, i.e. becoming heavier for increasing energies above the ankle \cite{aa14}, and leads to a distribution of arrival directions also in qualitative agreement with observations. The high degree of isotropy observed even at the highest CR energies requires strong extragalactic magnetic fields (EGMF) to be present in the local neighborhood between the Cen~A source and us, with typical root mean square field strengths $B_{\rm rms}$ of a few tens of nG. The presence of these EGMFs implies that to a large extent the CRs should diffuse as they travel from this source to us, and even at the highest rigidities achieved, typically of $E/Z\simeq 5$\,EeV, they can be deflected by 20$^\circ$ or more, and the particles that may reach us after turning around during their diffusive propagation will lead to a more isotropic contribution to the distribution of arrival directions. At the same time, the CR density should get enhanced in the local region as a consequence of the diffusion process. The energy dependence of the diffusion will significantly affect the observed spectrum, and accounting for the expected finite lifetime of the source also leads to a suppression of the spectrum at low rigidities \cite{be90,mo19b}. All these features are crucial to reproduce the different observations.

The hadronic interaction models used to interpret the air shower measurements, and hence to infer the masses of the CRs reaching Earth, have recently been significantly improved with the introduction of the EPOS.LHC-R model \cite{pi25}. Besides being constrained with a larger amount of updated experimental LHC results that include those of the recent runs of p-p collisions up to 13~TeV, the model has a comprehensive approach aiming to reproduce extensive data ranging from $e^+e^-$ colliders up to heavy-ion collisions, such as p-Pb or Pb-Pb ones. In addition to the hadronization through string fragmentation taking place in the `corona' of the interaction region, the inclusion of the collective hadronization in the `core' region, as well as hadronic rescattering of the final states, are key to reproduce the details of several data. The enhanced production of $\rho^0$ mesons, which decay to charged pions but not $\pi^0$, is also quite relevant as it leads to an enhanced production of muons in the air showers. The tuning to new experimental data also leads to larger elasticities in the hadronic interactions, giving rise to deeper predicted values for the depth of the maximum development of the air shower $X_{\rm max}$. As a consequence, the CR masses inferred from the measured values of $\langle X_{\rm max}\rangle$ turn out to be heavier, further enhancing the expected muon fluxes at ground level and alleviating the tensions that were present when previous hadronic interaction models were adopted \cite{ab24a} and solving to a large extent the muon deficit problem. 

As we show in the present work, the heavier inferred composition has important effects on the features of the UHECR scenario based on Centaurus~A as the dominant source above the ankle. In particular, we will find that the amount of light elements (H and He) required to be emitted from the nearby source gets reduced, which also affects the magnetic field-induced average deflections and hence the localized anisotropies as well as the predicted dipole component. We will also find that the features in the variance of the mass distribution can be better reproduced with the results obtained on the basis of this new hadronic model, and that in general the fit to the spectrum and composition is improved. We also present in the Appendix a new method to derive the mass information from the $X_{\rm max}$ measurements, which allows us to obtain it reliably over a wide energy range and with an improved accuracy  with respect to previous methods.

\section{Main characteristics of the Centaurus~A scenario}

We briefly describe here the main components that contribute to the observed CR fluxes in the scenario of ref.\,\cite{mo24}. The UHECRs above the ankle energy are mostly due to the Centaurus~A radiogalaxy, while a population of extragalactic sources distributed over space dominates the fluxes below the energy-spectrum ankle and may eventually provide a subdominant contribution at higher energies. The sources are assumed to emit the representative elements $j={\rm H}$, He, N, Si and Fe,\footnote{For simplicity, we take here N as the representative of the CNO group, although considering both C and O can slightly improve the fits \protect\cite{mo24}.} which are 
being emitted with a power-law spectrum having a rigidity dependent cutoff,\footnote{We just refer to the quantity $E/Z$ as the CR rigidity, measured in EeV.} so that
\begin{equation}
     \frac{{\rm d}\Phi_\alpha}{{\rm d}E}=\phi_0^\alpha\sum_jf_j^\alpha \left(\frac{E}{E_0}\right)^{-\gamma_\alpha}\, {\rm sech}\left(\frac{E}{Z_jR_{\rm cut}^\alpha}\right)^\Delta,
 \end{equation}
 where the index $\alpha$ identifies the population considered,  with $\alpha={\rm L}$ for the  low-energy  population or $\alpha=s$  for the nearby source. The elemental fractions are $f_j^\alpha$ and  $\phi_0^\alpha$ are the flux normalizations at a reference energy $E_0$ smaller than the cutoff rigidity $R_{\rm cut}^\alpha$. The parameter $\Delta$ characterizes the steepness of the cutoff, and 
 we will illustrate the results for $\Delta=1$ and 2.
 For definiteness, we will consider that the low-energy population has no source cosmological evolution, so that $\phi_0^{\rm L}$ is constant in time (an evolution according to, for instance, the star formation rate just leads to slightly lower inferred values of $\gamma_{\rm L}$). Other relevant parameters are related to the extra-galactic magnetic field required to be present in the local neighborhood, in the region containing Cen~A and our galaxy. We consider a homogeneous turbulent magnetic field, characterized by its root mean square strength $B_{\rm rms}$, its coherence length $l_{\rm coh}$ (that we take to be 100~kpc) and we assume that the turbulence has a Kolmogorov spectrum. This field will have the effect of isotropizing to a significant extent the distribution of CR arrival directions from the source. It will also lead to a modulation of the spectrum as a consequence of the diffusion process, which depends on the critical rigidity $R_{\rm c}\equiv eB_{\rm rms}l_{\rm coh}\simeq 0.9 (B_{\rm rms}/{\rm nG})(l_{\rm coh}/{\rm Mpc})$\,EeV. The modulation of the spectrum at low rigidities also depends on the lifetime of the source, since for finite lifetimes the low rigidity particles will not be able to reach the observer. We assume for definiteness that the source has been emitting steadily since a time $t_i \simeq (20\,{\rm to}\, 50)r_{\rm s}/c$, with $r_{\rm s}\simeq 4$\,Mpc being the source distance. The detailed expressions for the energy-dependent source modulation are given in ref.\,\cite{mo24} (see also refs.\,\cite{mo19b,mo22}).
 Another relevant factor is the Galactic magnetic field (GMF), that can lead to coherent deflections which in particular displace the direction of the dipolar anisotropy  and also slightly affect its amplitude. We consider for it the Jansson and Farrar model (JF12) \cite{ja12}, neglecting its random component given that the turbulent EGMF has significantly stronger effects in our scenario. 

 In addition to the previous extragalactic sources, we also include a Galactic contribution, taken from ref.\,\cite{mo19}, which contributes at the $\sim 10$\% level at 1\,EeV, being mostly composed of Fe nuclei at these energies. It gets strongly suppressed at higher energies. 

\section{Results and discussion}

We will fit the spectrum and composition data from the Pierre Auger Observatory, which dominates the cosmic-ray statistics at present, considering energies above $10^{17.8}\,{\rm eV}\simeq 0.63$\,EeV. For the spectrum we use the results from  \cite{ab21} and for the average and dispersion of the $X_{\rm max}$ distribution we use the values in \cite{yu19}, which were obtained for bins in lg$E$ of width 0.1. We obtain from them the values of $\langle\ln A\rangle$ and Var(ln$A$) using the method introduced in the Appendix, which generalizes the one used previously in \cite{ab13, yu19} and allows to derive these quantities more accurately.  For fits to the spectrum and composition information in other astrophysical scenarios, see for instance \cite{aa17b,ab23,ab24,ab24c}.

The model parameters to be fitted for the scenario introduced in the previous section are
 the elemental fractions $f^\alpha_j$ emitted at the sources by each population (which add up to one, so that only 4 are independent for each population), the associated spectral indices $\gamma_\alpha$, the two cutoff rigidities $R_{\rm cut}^\alpha$, the critical rigidity associated with the EGMF $R_{\rm c}$ as well as the time of emission considered $t_i$.\footnote{Also the flux normalization of the Cen~A source and of the low-energy population have to be determined, but this can be done analytically.} These last two quantities have a strong effect on the shape of the spectrum observed at Earth, as well as on the expected  anisotropies. In particular, the amplitude of the dipolar modulation at Earth scales approximately as the inverse of the emission time \cite{mo19}, and we will choose this time so as to approximately reproduce the observed dipole amplitude. The critical rigidity will instead be determined from the fit of the spectrum and composition data. Since we consider $l_{\rm coh}=100$\,kpc, the value of $R_{\rm c}$ will determine the required size of  $B_{\rm rms}\simeq 11\,{\rm nG}(R_{\rm c}/{\rm EeV})(100\,{\rm kpc}/l_{\rm coh})$.

\begin{table}[t]
\addtolength{\tabcolsep}{-1pt}
\centering

\scalebox{0.87}{ 
\begin{tabular}[H]{ @{}c| c  c c c c c c | c c c c c c c|c|c| c @{}}
\multicolumn{17}{c}{ model G-NE-4Mpc,  EGMF with $l_{\rm coh}=100$~kpc }\\
\hline
$\Delta$ & $\gamma_{\rm s}$ & $R_{\rm cut}^{\rm s}$ & $f^{\rm s}_{\rm H}$ & $f^{\rm s}_{\rm He}$ & $f^{\rm s}_{\rm N}$  & $f^{\rm s}_{\rm Si}$ & $f^{\rm s}_{\rm Fe}$ &$\gamma_{\rm L}$ & $R_{\rm cut}^{\rm L}$ & $f^{\rm L}_{\rm H}$ & $f^{\rm L}_{\rm He}$ & $f^{\rm L}_{\rm N}$& $f^{\rm L}_{\rm Si}$& $f^{\rm L}_{\rm Fe}$ & $R_{\rm c}$ & $\frac{ct_i}{r_s}$ & $\frac{\chi^2}{\rm dof}$  \bigstrut[t]\\ 
 & &  [EeV] & & &  [\%] &  &  &  & [EeV]  &  & & [\%] & & &  [EeV] & &\bigstrut[b]\\
\hline
1& 0.62 & 1.1 & 87.9 & 7.8 & 4.5 & 0.6 & 0.2 & 3.2 & 48 & 32.4 & 0.1 & 51.9 & 14.0 & 1.7 &  2.5 & 40 & 0.86 \bigstrut[t]\\
2 & 2.5 & 3.2 & 20.4 & 1.9 & 38.6 & 18.8 & 20.3 &  3.1 & 53 & 31.0 & 0 & 59.5 & 5.1 & 4.4 & 2.8 & 25 & 0.98\bigstrut[t]\\
\hline
\multicolumn{17}{c}{ idem with $E\times 1.14$} \bigstrut[t]\\
\hline
1& 0.82 & 1.4 & 77.6 & 15.3 & 6.1 & 0.8 & 0.3 &  3.3 & 143 & 30.5 & 0 & 59.9 & 0 & 9.6 & 2.8 & 40 & 1.03\\
2& 2.5 & 4.0 & 20.2 & 7.8 & 39.0 & 14.6 & 18.4  & 3.2 & 94 & 29.7 & 0 & 61.3 & 0.2 & 8.8 & 3.1 & 25 & 1.16\\
\hline
\end{tabular}}
\label{table:1}

\caption{Parameters of the fit to the flux and composition for the different scenarios including an EGMF with $l_{\rm coh}=100$~kpc and a source emitting steadily since an initial time $t_i$. Results are shown for $\Delta=1$ and 2, and also when rescaling the observed energies by a factor 1.14. }
\label{tab1}
\end{table}

\begin{figure}[t]
    \centering
    \includegraphics[width=0.45\textwidth]{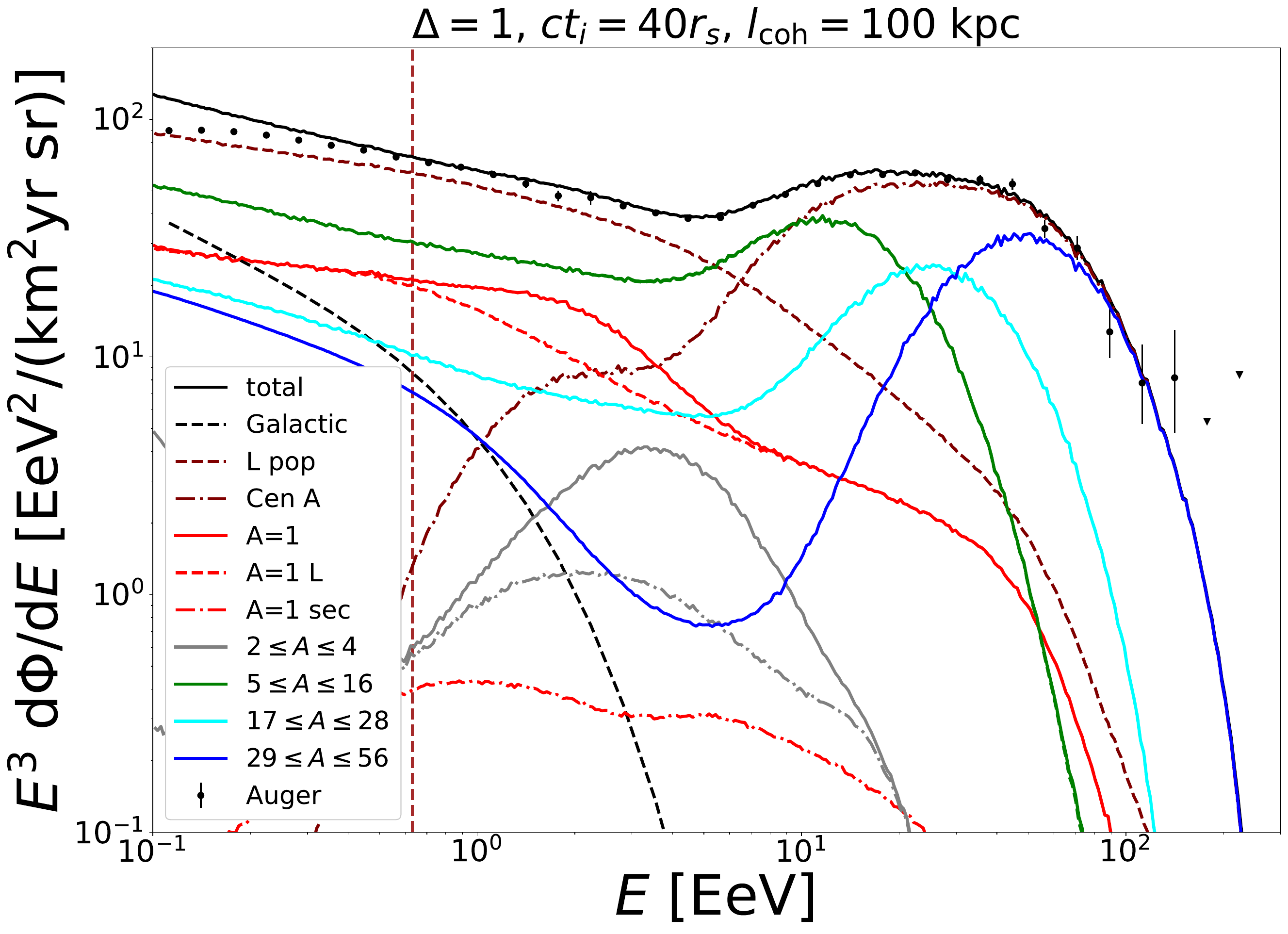}\includegraphics[width=0.45\textwidth]{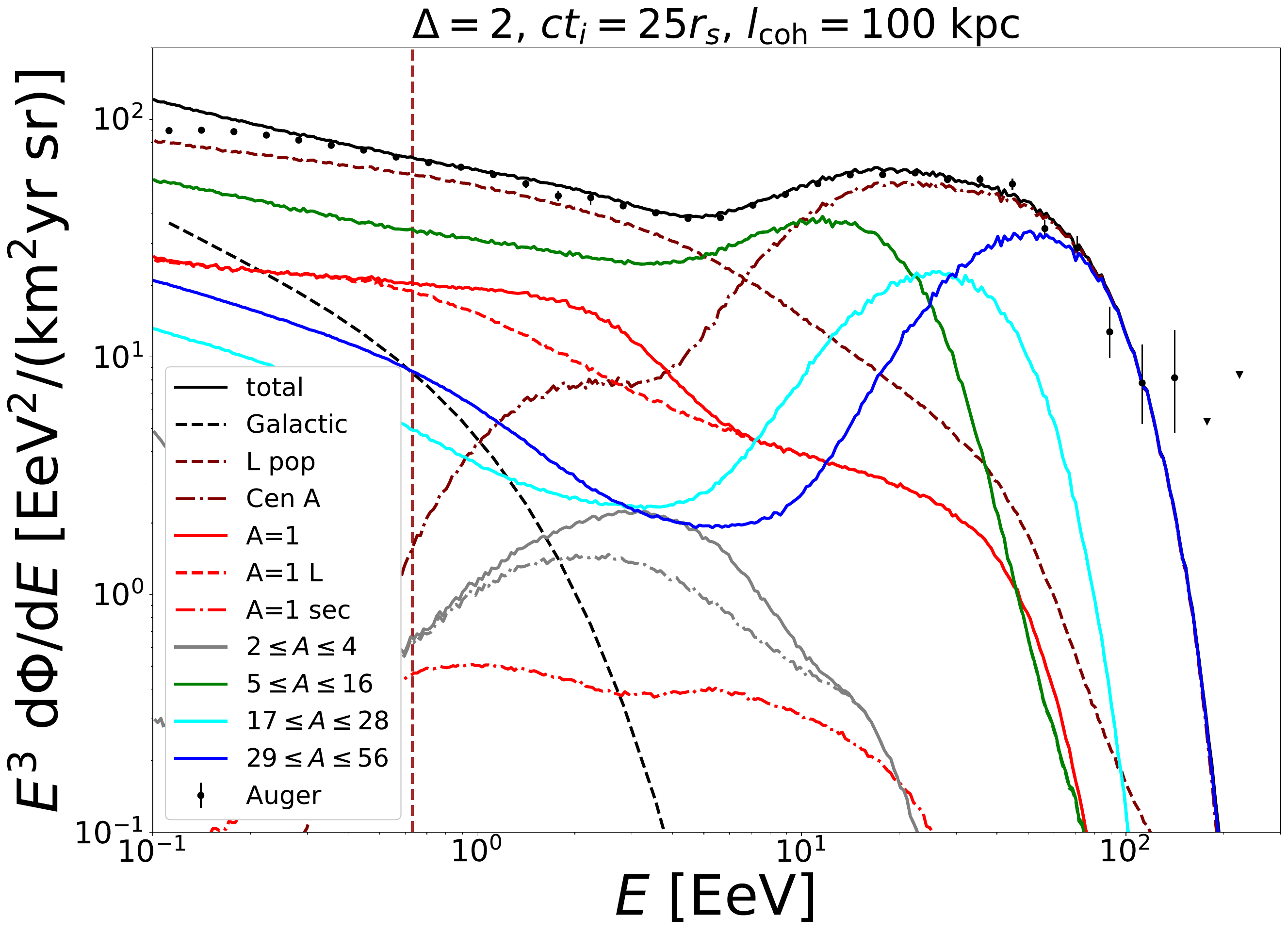}\\
    \includegraphics[width=0.45\textwidth]{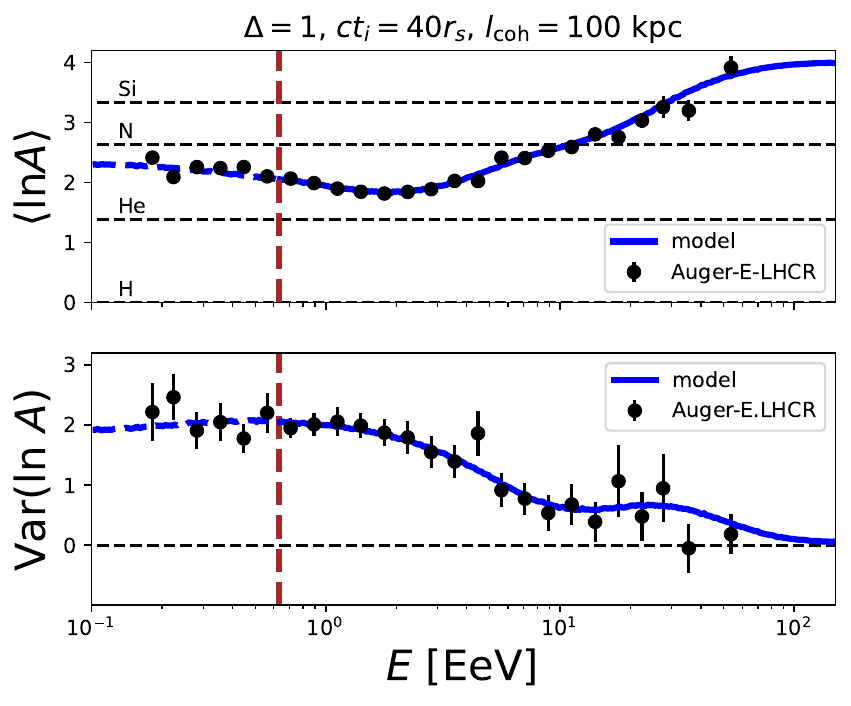}\includegraphics[width=0.45\textwidth]{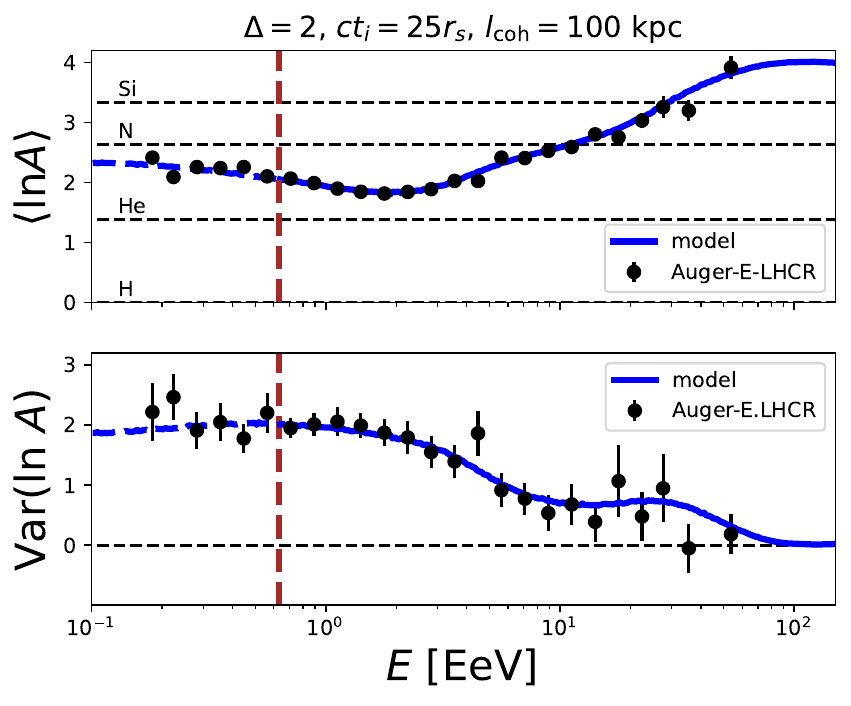}\\
   
    \caption{Spectrum (top) and composition (bottom) obtained from the fit, compared to the measured data (dots). Results for two different values of the cutoff parameter are shown (left for $\Delta=1$ and right for $\Delta=2$). In the spectrum plots the different components are shown. Vertical dashed lines indicate the threshold of the fit ($10^{17.8}$\,eV).}
    \label{SCnoshift}
\end{figure}

We show in Fig.~\ref{SCnoshift} the resulting fit to the data for the different cases considered, reporting in Table~1 the parameters obtained in the fit for the different values of $\Delta$. One finds that in all cases the agreement between the model predictions and the data is quite good, with $\chi^2/{\rm dof}$ being close to one.  

One can see that the composition above the ankle, where the CRs are mostly due to the Cen~A source, is predominantly given by N, Si and Fe, which appear as a sequence of bumps with increasing masses for increasing energies. The fact that now the Fe nuclei (rather than the CNO ones) dominate the spectrum near the steep suppression present above 50~EeV implies that the required source cutoff should be of only few EeV. This has the effect of significantly suppressing the CNO nuclei above around 20~EeV, and this component turns out to be responsible for the instep spectral feature observed near 15\,EeV.\footnote{Changes of this kind in the spectral and composition features were also found with previous hadronic models when considering systematic ad-hoc shifts on the energy or $X_{\rm max}$ scales \cite{ab24c},\cite{mo24},\cite{vi25}.} Contrary to what happened with previous hadronic models, now the contribution of the light H and He components from the nearby source is almost negligible above 10~EeV. The protons still contribute sizeably at few EeV, being responsible for the minimum in $\langle \ln A\rangle$ observed near 3\,EeV, while the He contribution is subdominant.
 The low source cutoff value allows us to obtain quite a good fit to the observed features of $\langle\ln A\rangle$ and Var(ln$A$), something which was not possible with the previous hadronic models. In particular, the fact that N largely dominates near 10\,EeV leads to a strong reduction in Var(ln$A$) around this energy, and at 30\,EeV there is a significant mix of N, Si and Fe elements which leads to a slight enhancement of the variance, which then becomes smaller as the Fe nuclei become the dominant element. 
Note that the inferred source spectral index of the nearby source depends sensitively on the assumed shape of the cutoff function, with $\Delta=1$ requiring much harder spectra (and lower cutoff rigidities) than in the case of $\Delta=2$. Given that the source spectra are softer for $\Delta=2$ than for $\Delta=1$, this also implies that the required value of $ct_i$ should be smaller for larger $\Delta$, so that the enhanced suppression at low rigidities gives rise to a similar spectrum at Earth for the different elements.
The required critical rigidity $R_{\rm c}$ turns out to be a few EeV, and let us mention that its value is strongly correlated with the value of $\gamma_s$, with small correlated changes in them leaving the $\chi^2$ almost unchanged.
 
Regarding the low-energy population, it  consists predominantly of N, H and eventually some Si or Fe nuclei, and actually a large value of its cutoff, of order 50 to 100~EeV, is preferred. This leads to the subdominant H component extending up to the suppression region, helping to account for the small bump in the spectrum observed near 50~EeV. Also note that the highest energy nuclei are affected by photodisintegrations during their journey and lead to a small amount of secondary H and He nuclei present at few EeV, which were displayed separately in the plot with dot-dashed lines.  Since the low-energy population has a steep spectrum and provides a small contribution at the highest energies, its associated spectral index does not depend significantly on the assumed value of the cutoff shape parameter $\Delta$. We have also found that adopting a stronger source evolution for this component than the non-evolving case considered here would lead to slightly smaller inferred values of $\gamma_{\rm L}$, given that propagation effects tend to steepen the spectrum at Earth in the presence of strong evolution. 

The Galactic population provides a small contribution to the spectrum at the lowest energies considered in the fit, but being mostly Fe at these energies it helps to account for the values of Var(ln$A$) determined near 1\,EeV. 
The vertical dashed line in the plots indicates the threshold of the fit. At lower energies, which are not included in the fit, one may see that the predicted spectrum appears to overshoot the data points, so that some additional ingredients may need to be considered to account also for those observations. These could be related to the presence of a magnetic horizon due to the finite source density of the low-energy population \cite{le05}, a change in the power-law slope of the spectrum emitted at the sources or eventually a different source evolution for this population.

\begin{figure}[t]
    \centering
    \includegraphics[width=0.45\textwidth]{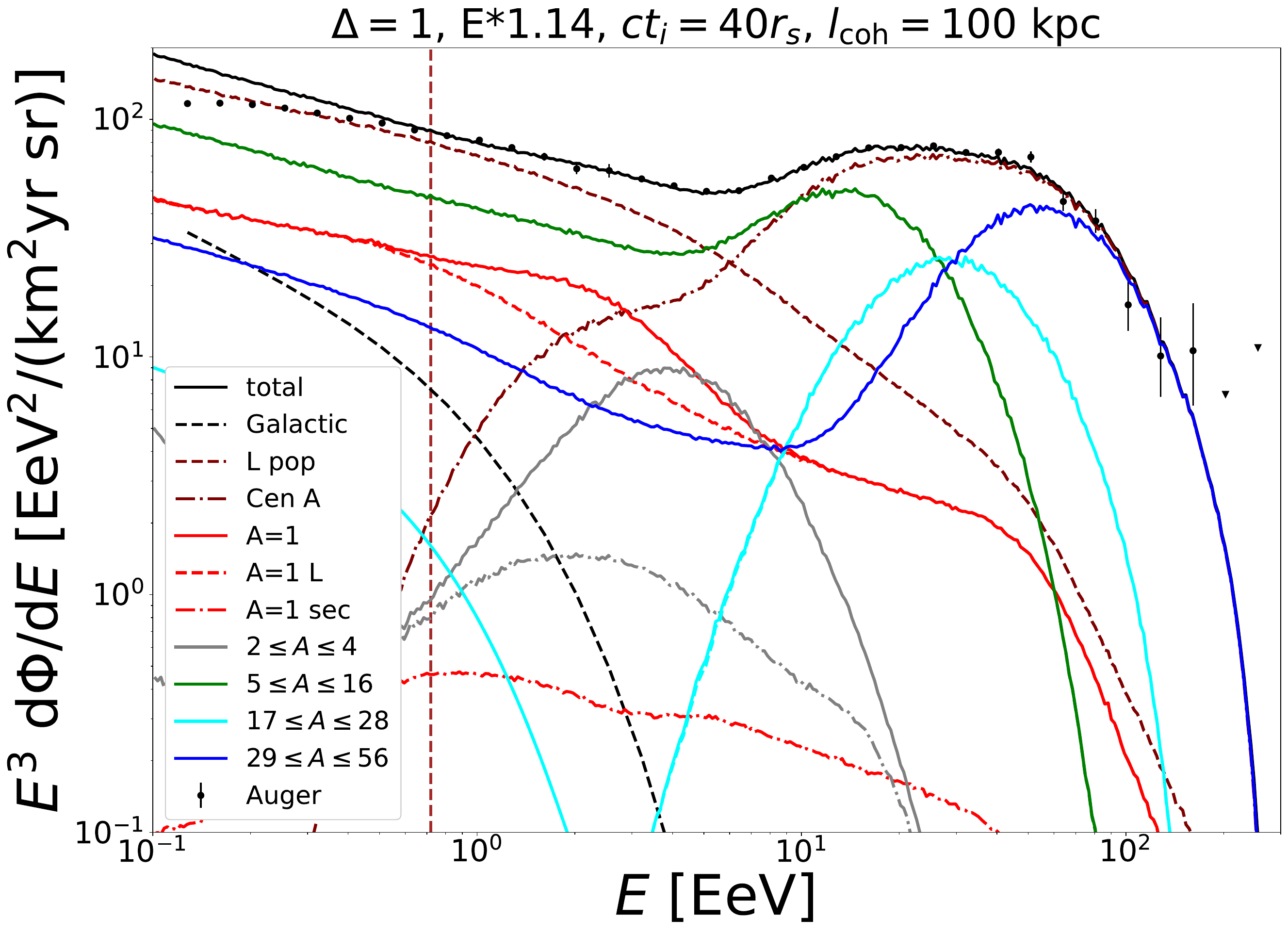}\includegraphics[width=0.45\textwidth]{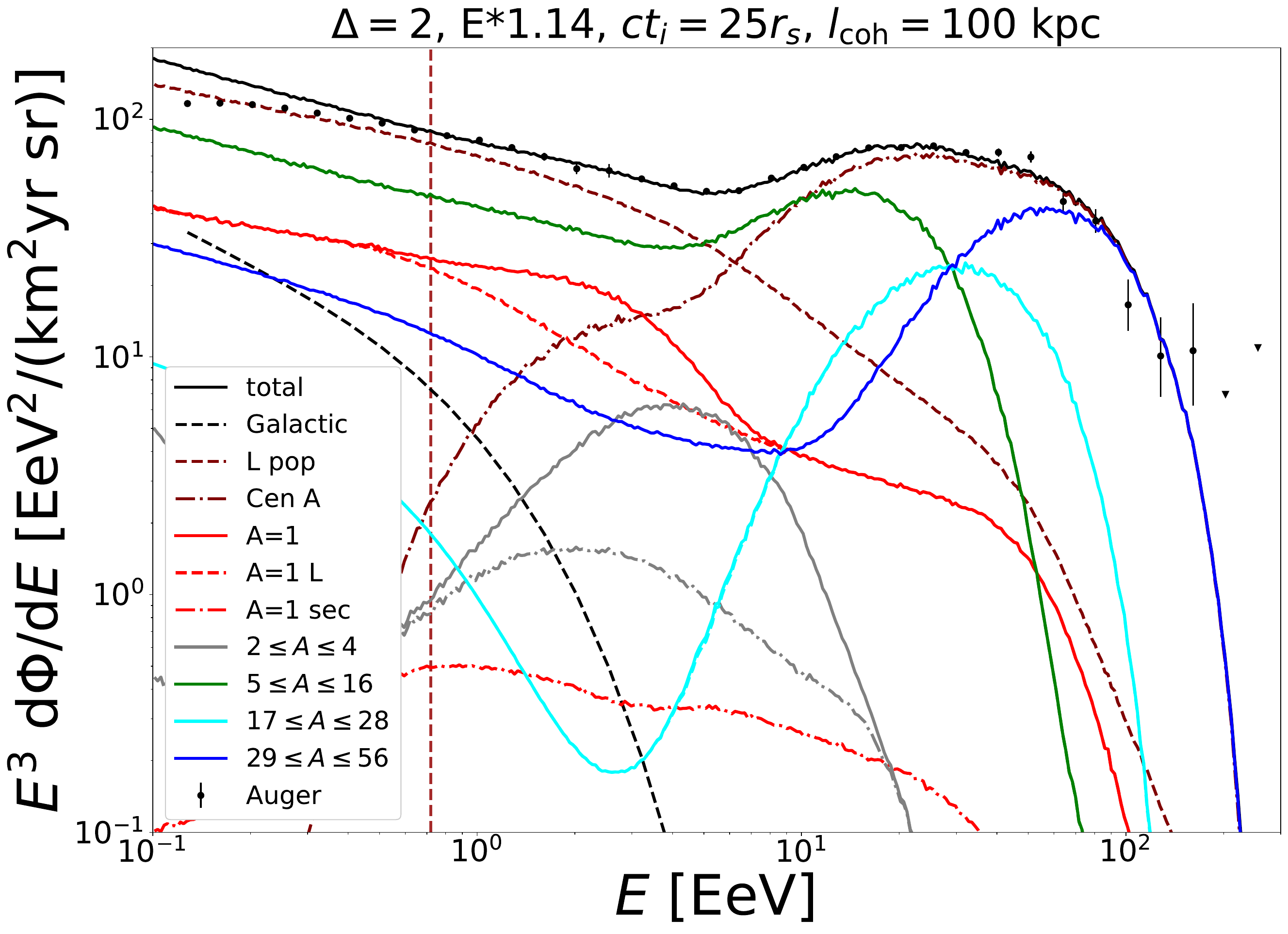}\\
    \includegraphics[width=0.45\textwidth]{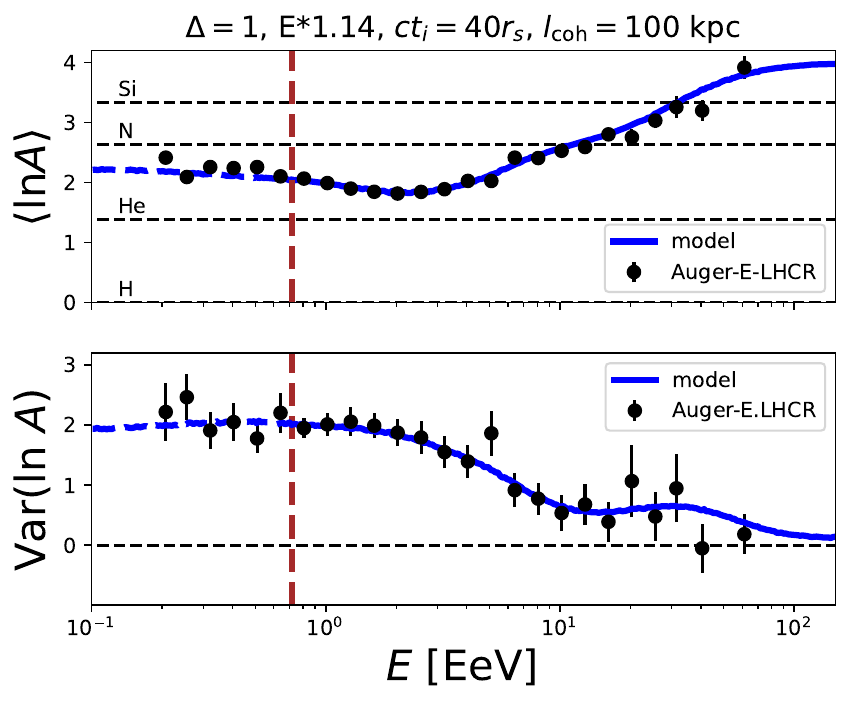}\includegraphics[width=0.45\textwidth]{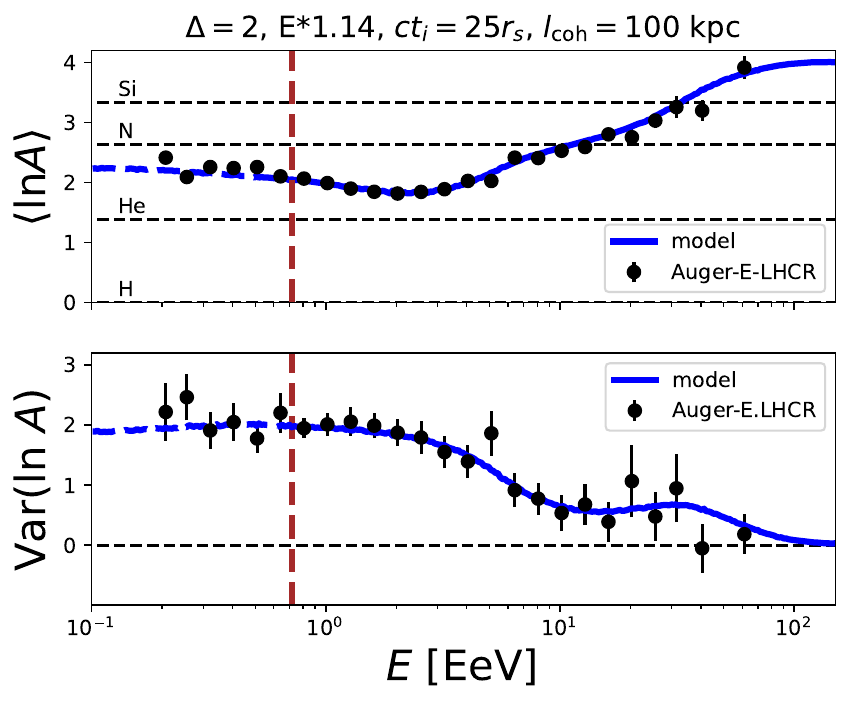}\\
   
    \caption{Same as in Fig.~\ref{SCnoshift} but shifting up the measured energies by a factor of 1.14.}
    \label{SCshift1}
\end{figure}

A relevant systematic effect in these analyses is related to the uncertainty in the energy scale of the measured cosmic rays, which for the Auger experiment  amounts to about $\sigma_E({\rm syst})\simeq 14$\% \cite{ab21}. 
We have found that increasing the energy scale by 1\,$\sigma_E({\rm syst})$ can slightly affect the qualitative features obtained.   The corresponding results are shown in Fig.~\ref{SCshift1}. One can note that with the positive energy shift the Fe contribution gets somewhat enhanced above 30\,EeV. Also, as is apparent from the table, the required values of the cutoff rigidities are larger. The opposite would result if negative shifts were considered.

\begin{figure}[t]
    \centering
    \includegraphics[width=0.6\textwidth]{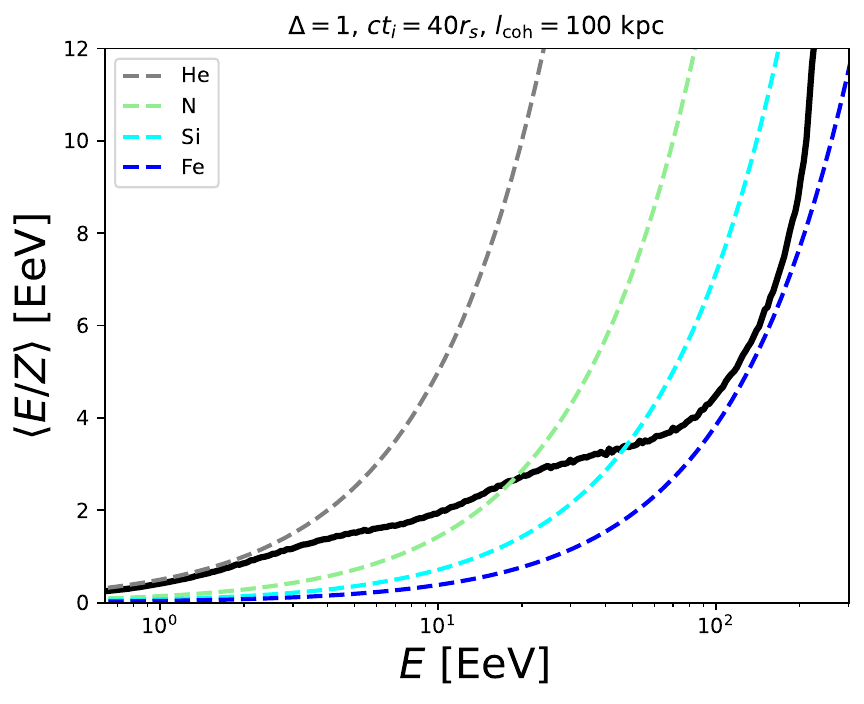}
   
    \caption{Average rigidity as a function of energy for the scenario with $\Delta=1$ and $ct_i=40r_s$. For reference, the expected rigidity for the different elements are also indicated (dashed lines)}
    \label{rigidity}
\end{figure}

Given that the composition of the cosmic rays becomes increasingly heavy, one finds that the associated mean rigidities of the observed cosmic rays are moderate, not exceeding 4~ EeV at energies below 100~ EeV. Their dependence with energy is illustrated in Fig.~\ref{rigidity} for the model with $\Delta=1$ and no energy shift, and similar results are also obtained in the other cases.

\begin{figure}[t]
    \centering
    \includegraphics[width=0.45\textwidth]{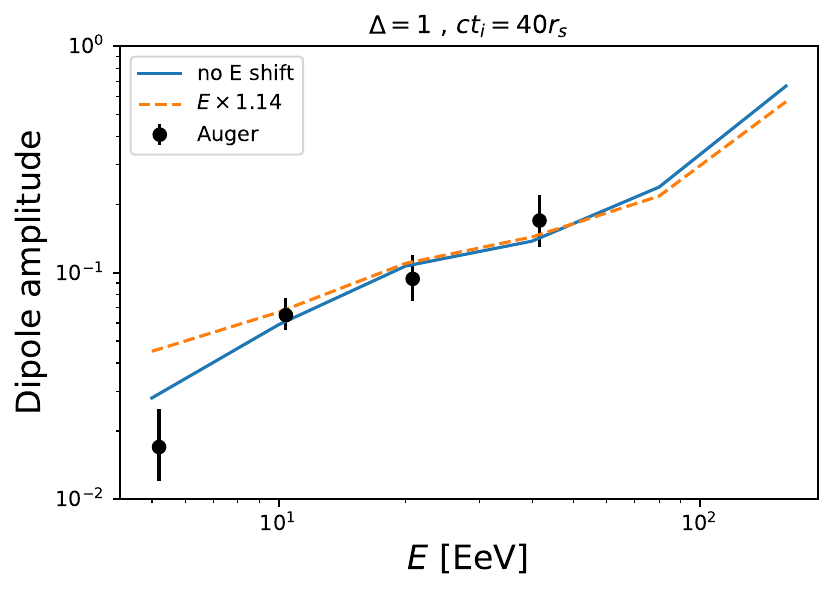}\includegraphics[width=0.45\textwidth]{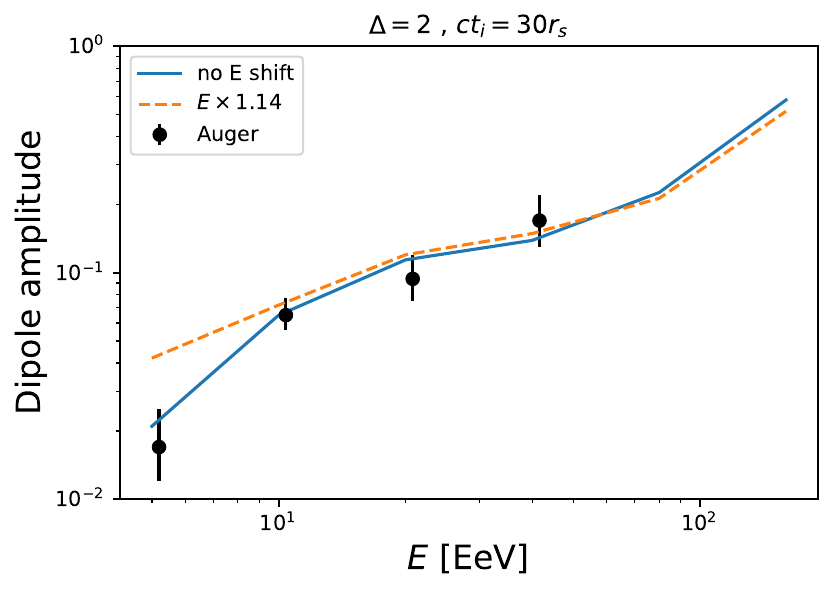}\\

    \caption{Dipole amplitude vs. energy for the scenarios with $\Delta=1$ (left) and $\Delta=2$ (right) both for the nominal energies and after shifting the measured energies by a factor of 1.14 and for the parameters shown in the Table.}
    \label{dips}
\end{figure}

Another important observable is the dipolar anisotropy, and in Fig.~\ref{dips} we show the resulting total dipolar amplitudes for the scenarios with $\Delta=1$ (left panel) and $\Delta=2$ (right panel), together with the measurements from \cite{ab24b}. Solid lines are for the cases without energy shifts and dashed lines for those with energies shifted  by +1$\sigma_E({\rm syst})$. One can see that, with the appropriate choice of $t_i$ (equal to $40r_s/c$ for $\Delta=1$ and $25r_s/c$ for $\Delta=2$), the results are in good agreement with the data, both in the overall amplitude as well as in the tendency towards an increasing dipole amplitude with increasing energy. In this scenario, the increasing dipole amplitude is directly related to the increasing average rigidity that was shown in Fig.~\ref{rigidity}. Note that the associated emission time of several tens of $r_s/c$ corresponds to CR travel times of up to several hundred million years, which is within the expected range for the Cen~A activity. The slight overshoot in the dipole amplitude present in some cases in the lowest energy bin (between 4 and 8\,EeV), could be alleviated for instance if at those low energies the CR  emission time were larger than assumed, or if the Galactic component were to contribute to the dipole in the opposite direction than the Cen~A source. Note that even if the Galactic contribution to the flux may be quite small at those energies, its intrinsic anisotropy is expected to be large, so that its contribution may not be negligible. Also the low-energy source population, which was assumed here to be isotropic, may actually have a non-negligible contribution to the dipole below  the ankle energy.

\begin{figure}[t]
    \centering
    \includegraphics[width=0.48\textwidth]{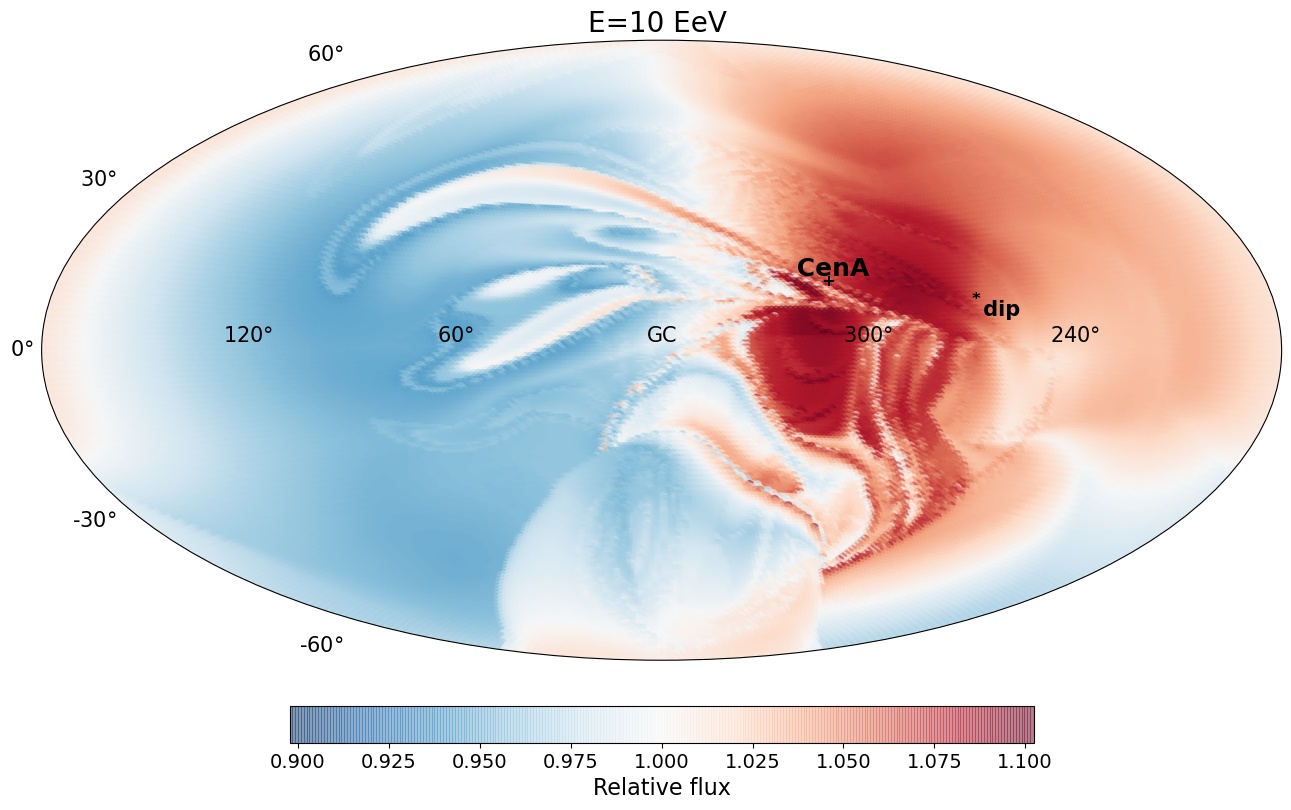}\includegraphics[width=0.48\textwidth]{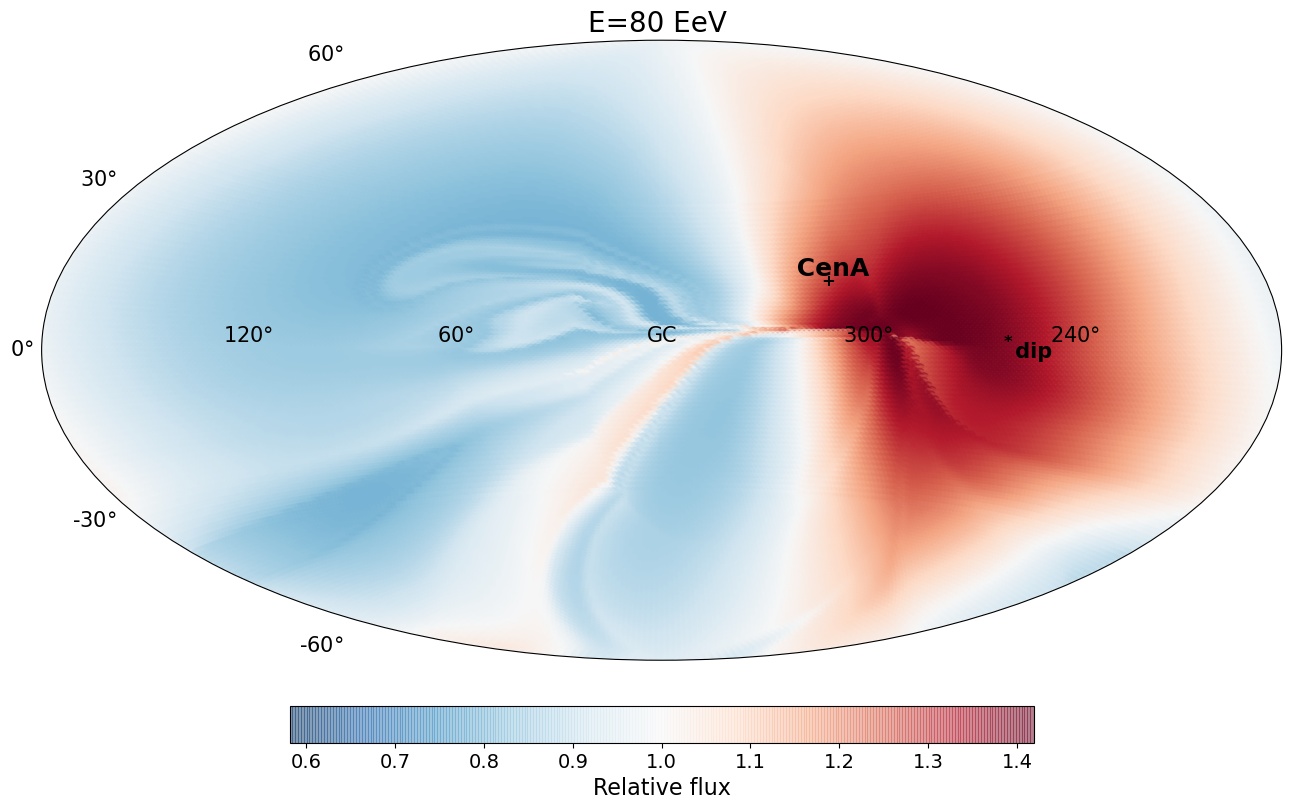}\\

    \caption{Maps in Galactic coordinates of the relative flux for the scenario with $\Delta=1$ and no energy shift. Left panel is for $E=10$~EeV while the right panel for $E=80$~EeV.}
    \label{maps}
\end{figure}

To better illustrate the expected arrival direction
distribution, we show in Fig.\,\ref{maps} the map of the relative fluxes for the particular case with $\Delta=1$ and no energy shift, which provides the best fit to the spectrum and composition data. One can see that the features are in very good qualitative agreement with  those observed \cite{go23b}, with a growing dipolar component and also some more localized  excesses showing up near the Cen\,A region. The reconstructed dipole direction (indicated as `dip' in the maps) appears also displaced with respect to the Cen~A position in the direction towards the outer galactic spiral arm. This is due to the  effects of the coherent GMF deflections. The more localized contribution appearing  closer to the Cen~A direction is due to the highest rigidity (lighter) particles, which suffer smaller deflections. Let us note that the fact that in this scenario the light H and He contributions from the source are very suppressed above the ankle energy implies that no strong localized anisotropy is expected around 10 to 20~EeV around the Cen~A location due to these elements, which would otherwise be analogous to the one produced by the CNO group for energies around 30 to 40~EeV. This could provide a sensitive test of the predictions of this scenario. On the other hand, the precise direction of the dipole and of the more localized flux excesses should depend on the details of the GMF as well as on the eventual contribution from a non-random component of the EGMF.

\section{Conclusions}
We have found that the new EPOS.LHC-R hadronic model has important implications for the the UHECR scenarios, in particular for that in which the Centaurus~A radiogalaxy is the main contributor to the CRs observed above the ankle. This new hadronic model leads to a heavier inferred composition, which in the scenario considered here can be achieved by means of a reduced value of the source cutoff rigidities. Now the suppression of the spectrum present around 50~EeV is mostly due to Si and Fe nuclei, rather than being due to the CNO component as was the case with previous hadronic models. In turn, the CNO component is now mainly responsible for the instep spectral feature. The smaller rigidities associated with the heavier inferred composition also affect the arrival directions, and one can naturally reproduce the main features of the observed dipolar distribution with reasonable assumptions on the emission history of the source and adopting EGMF strengths of a few tens of nG. In addition, more localized anisotropies appear near the Centaurus~A direction due to the highest rigidity particles observed. The suppression of the He nuclei from the source above the ankle energy implies that the localized anisotropies due to He nuclei  in the 10 to 20 EeV range, analogous to those observed above 30~EeV due to the CNO group elements, should be expected to be suppressed. 
The accuracy with which the observations can be accounted for has improved significantly with this new hadronic model, and the main features of the measurements can be better understood within this scenario.

\section*{Acknowledgments}
We are grateful to Tanguy Pierog for providing the $X_{\rm max}$ moments from the EPOS.LHC-R hadronic-model simulations.

\section*{Appendix}

In order to infer the cosmic ray composition from the measured distribution of the depths of shower maxima $X_{\rm max}$, the usual approach is to compare the average $X_{\rm max}$ and its dispersion $\sigma(X_{\rm max})=\sqrt{{\rm Var}(X_{\rm max})}$ with the expectations obtained from simulated showers produced by CRs with different mass numbers $A$. These simulations depend on the hadronic interaction model adopted to describe the CR interactions with the air molecules, and hence the inferred values for the relevant quantities, which are $\langle\ln A\rangle$ and Var(ln$A$) for a given dataset, also depend on the hadronic model considered.

We present here a new approach to perform this task and to extract the mass information, which generalizes the one used  by the Auger  Collaboration in refs.~\cite{ab13, yu19}. It adopts physically motivated  parameterizations of the values of $\langle X_{\rm max}\rangle$ and Var($X_{\rm max}$) obtained with simulated showers. In particular, we exploit an important basic property of the average depth of shower maximum of different nuclei which results from the so called superposition picture, in which the shower due to a nucleus of mass number $A$ can be considered as being the superposition of $A$ nucleon  showers of energy $E/A$. This implies that
\begin{equation}
    \bar{X}_{\rm max}^A(E)\simeq \bar{X}_{\rm max}^p(E/A),
\end{equation}
 where  the overline denotes the average over simulations with a fixed mass and we will use $\langle X_{\rm max}\rangle$ instead to denote the average over the data set, which will generally contain CRs with several different masses.

 The average depth of shower maximum of proton showers is usually expressed as 
\begin{equation}
    \bar{X}_{\rm max}^p(E)= X_0+D\ {\rm lg}E,
\end{equation}
with lg$E\equiv {\rm log}_{10}(E/{\rm EeV})$. Here
 $D$ is the elongation rate, i.e. the change of $\langle X_{\rm max}\rangle$ per decade of energy. Previous methods~\cite{ab13, yu19} considered this elongation rate to be a constant. However, simulations show that for a single element the elongation rate is actually energy dependent and that it should slightly decrease for increasing energies. We will hence consider  (see also ref.~\cite{mo19})
\begin{equation}
    D(E)=D_0+D_1 {\rm lg}E.
\end{equation}

Using the previous two equations we will then adopt for nuclei of mass number $A$ the expression
\begin{equation}
    \bar{X}_{\rm max}^A(E)= X_0+[D_0+D_1 {\rm lg}(E/A)]\,{\rm lg}(E/A).
    \label{xamxpar}
\end{equation}
 The behavior of the different $X^A_{\rm max}$ is then parameterized with just the three parameters $X_0$, $D_0$ and $D_1$, and the superposition assumption of Eq.~(\ref{xamxpar}) will naturally reproduce the mass dependence of $\bar{X}_{\rm max}$. 
Since simulations for different nuclei are usually available in the same energy range, we will determine the coefficients by fitting the values of $\bar{X}_{\rm max}^{p}$ and  $\bar{X}_{\rm max}^{Fe}$ so as to also cover a wider range of values of $E/A$.

\begin{figure}[t]
    \centering
    \includegraphics[width=0.49\textwidth]{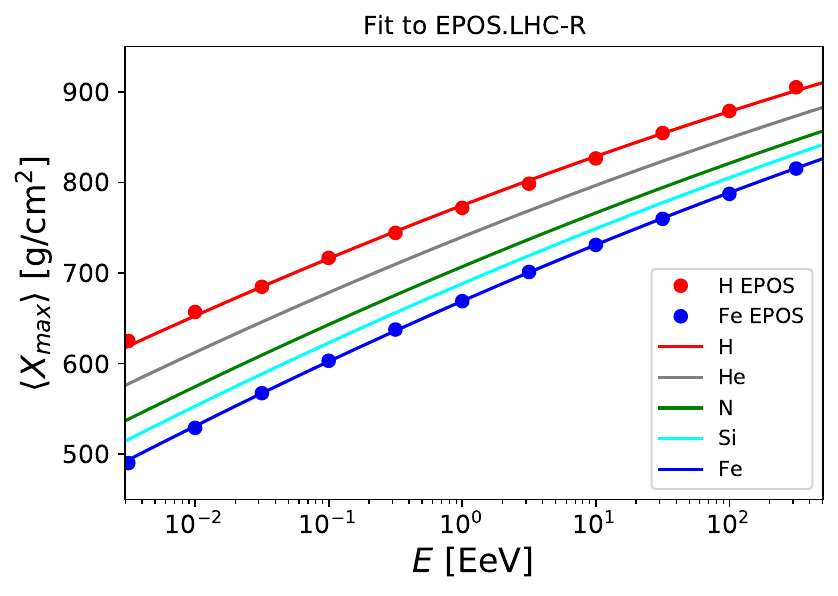}    \includegraphics[width=0.49\textwidth]{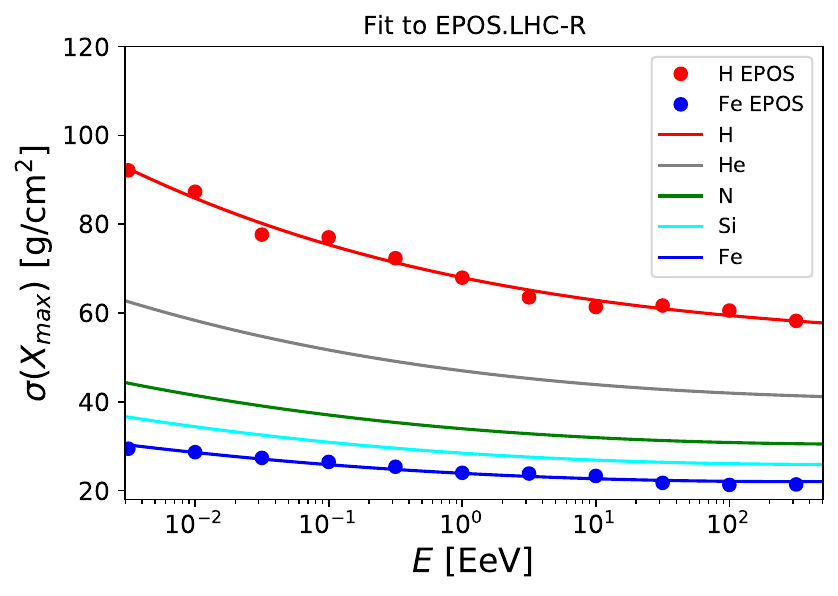}\\    
   
    \caption{Values of $\langle X_{\rm max}\rangle$ (left) and $\sigma(X_{\rm max})$ (right) for the EPOS.LHC-R simulations of H and Fe from \cite{pi25} (dots). Also shown with lines are the fits obtained and the expectations for He, N and Si.}
    \label{fig1}
\end{figure}

We show in Fig.~\ref{fig1} (left panel) the results of simulations with H and Fe primaries for the EPOS.LHC-R hadronic model from ref.~\cite{pi25}. We will fit the energy range between the spectral knee at 3\,PeV up to 300\,EeV, so as to cover the energies explored with air-shower experiments. The results of simulations are shown with dots in the figure. Also shown are the fits obtained with the previous parameterizations, which lead to $X_0=774.6$\,g/cm$^2$, $D_0=56.4$\,g/cm$^2$ and $D_1=-2.3$\,g/cm$^2$.  As one can see, the agreement is quite good in the whole energy range for both nuclei, and clearly the energy dependence of the elongation rate is responsible for the observed curvature of the different lines.  Also shown in the figure are the expectations for He, N and Si nuclei obtained with the approach just described. Let us note that previously available methods, which rely on a constant elongation rate, are not able to properly fit these observables  in such a wide range of energies. 

Regarding the $X_{\rm max}$ fluctuations, the simulations show that they are significantly larger than the naive expectation from the superposition model, which is $\sigma(X^A_{\rm max})\simeq \sigma(X^p_{\rm max})/\sqrt{A}$. This can be understood according to the so-called semi-superposition model \cite{en92} as resulting from the fact that in every interaction of a nucleus with an air molecule actually several nucleons may be wounded, besides some nuclear fragments being eventually emitted. This implies that the sub-showers that are produced until the point in which the original nucleus gets totally split are correlated, and hence the resulting shower cannot be considered as being a superposition of $A$ independent nucleon showers.  As a consequence,  the fluctuations in $X_{\rm max}$ for a heavy nucleus tend to be closer to those of lighter nuclei, being larger than the naive expectation by a factor that can be even greater than two for Fe induced showers. This behavior does not affect however the predictions for $\bar{X}_{\rm max}^A$, so that the relation in Eq.~(\ref{xamxpar}) is robust. 

We will then propose that due to the semi-superposition, the effective number of independent nucleon sub-showers can be expressed as $N_{\rm eff}\simeq A^\alpha$, where the parameter  $\alpha$ will turn out to be about 0.6. To better reproduce the results in the simulations for the whole energy range explored, we will actually consider that $\alpha=\alpha_0+\alpha_1\ {\rm lg}E$. We will hence adopt
\begin{equation}
    {\rm Var}(X^A_{\rm max})(E)=\frac{{\rm Var}(X^p_{\rm max})(E/A)}{A^\alpha}.
    \label{varxa}
\end{equation}
For the variance of the proton showers,  we will consider  the power-law shape
 \begin{equation}
     {\rm Var}(X_{\rm max}^p)=p_0+p_1 \exp(-p_2\,{\rm lg}E).
   \label{varxp}
   \end{equation}

 The fit to both the proton and Fe simulations of EPOS.LHC-R  leads to $p_0=2821\,{\rm g}^2/{\rm cm}^4$, $p_1=1799\,{\rm g}^2/{\rm cm}^4$, $p_2=0.46$, $\alpha_0=0.62$ and $\alpha_1=-0.035$. The results of the simulations and the corresponding fits are displayed in Fig.\,\ref{fig1} (right panel), with the agreement being quite good in the whole energy range. Also shown are the predictions for He, N and Si. 
 
In order to characterize the distribution of CR masses we need now to relate the $X_{\rm max}$ moments with those of ln$A$. Using that
\begin{equation}
    \bar{X}^A_{\rm max}(E)=\bar{X}^p_{\rm max}(E/A)=\bar{X}^p_{\rm max}(E)+x_1\ln A+x_2(\ln A)^2, 
\end{equation}
where 
\begin{align}
    x_1 &\equiv -[D_0+2D_1{\rm lg}E]/\ln 10 ,\nonumber  \\
    x_2 &\equiv D_1/(\ln 10)^2 ,
\end{align}
we then get for the average
\begin{equation}
    \langle X_{\rm max}\rangle=\bar{X}^p_{\rm max}+x_1\langle\ln A\rangle+x_2({\rm Var}(\ln A)+\langle\ln A\rangle^2).
\end{equation}
For the computation of the variance we will use the law of total variance, which states that 
\begin{equation}
    {\rm Var}(X) = E[{\rm Var}(X|Y)]+{\rm Var}(E[X|Y])
\end{equation}
and, in our case, taking $X=X_{\rm max}$ and $Y=\ln A$ it will read 
\begin{equation}
    {\rm Var}(X_{\rm max})=\langle {\rm Var}(X^A_{\rm max})\rangle + {\rm Var(\bar{X}^A_{\rm max})}.
    \label{totalvar}
\end{equation}
This just means that the variance of $X_{\rm max}$ is the sum of the average of the variances of each element (i.e. of the shower to shower fluctuations obtained in simulations) plus the variance of the averages $\bar{X}^A_{\rm max}$  obtained for each of the elements. 
 Something which turns out to be useful in some cases is to perform Taylor's expansions so that for any function $F(\ln A)$ one has that 
\begin{equation}
    \langle F(\ln A)\rangle \simeq F(\langle\ln A\rangle)+\frac{1}{2}F''(\langle\ln A\rangle){\rm Var}(\ln A),
\end{equation}
where
\begin{equation}
    F''(\ln A)\equiv \frac{{\rm d}^2 F}{{\rm d}\ln A^2}(\ln A).
\end{equation}
 We obtain in particular that
\begin{equation}
    \langle {\rm Var}(X^A_{\rm max})\rangle \simeq {\rm Var}(X^A_{\rm max})\left|_{\langle\ln A\rangle}\right.+\frac{1}{2}\left.{\rm Var}(X^A_{\rm max})''\right|_{\langle\ln A\rangle}{\rm Var}(\ln A).
\end{equation}
Writing
\begin{equation}
    {\rm Var}(X^A_{\rm max})(\ln A)= G(\ln A)/\exp(\alpha\ln A),
\end{equation}
where from Eqs.\,(\ref{varxa}) and (\ref{varxp})
\begin{equation}
    G(\ln A)\equiv p_0+p_1 \exp[-p_2\,({\rm lg}E-\ln A/\ln10)],
\end{equation}
we obtain after explicit computation that
\begin{equation}
    \langle {\rm Var}(X^A_{\rm max})\rangle \simeq\frac{ \left(G(\langle\ln A\rangle)+\frac{{\rm Var}(\ln A)}{2}\left[ \alpha^2G(\langle\ln A\rangle)-2\alpha G'(\langle\ln A\rangle)+G''(\langle\ln A\rangle)\right]\right)}{ \exp(\alpha\langle\ln A\rangle)},
    \label{avvarxa}
\end{equation}
with
\begin{align}
    G'(\ln A)&= \frac{p_1  p_2}{\ln 10}\exp[-p_2\,({\rm lg}E-\ln A/\ln10)] , \nonumber\\
    G''(\ln A)&= p_1\left(\frac{  p_2}{\ln 10}\right)^2\exp[-p_2\,({\rm lg}E-\ln A/\ln 10)]  .
\end{align}
We can also obtain that 
\begin{equation}
    {\rm Var}(\bar{X}^A_{\rm max})={\rm Var}(\ln A)\left[(x_1+2x_2\langle\ln A\rangle )^2-x_2^2{\rm Var}(\ln A)\right].
    \label{varavx}
\end{equation}
Combining Eqs.\,(\ref{avvarxa}) and (\ref{varavx}) by means of Eq.\,(\ref{totalvar}), the final expression for Var$(X_{\rm max})$ is then obtained.

It is clearly not possible to invert these equations to obtain a closed expression for  $\langle \ln A\rangle$ and Var(ln$A$). To obtain these quantities we will just consider, for each energy bin in which the measured values  
$\langle X_{\rm max}\rangle_d\pm \sigma_X$ and  ${\rm Var}(X_{\rm max})_d\pm \sigma_{\rm Var}$ are determined, the $\chi^2$ function
\begin{equation}
    \chi^2(\langle \ln A\rangle, {\rm Var}(\ln A))=\frac{\left(\langle X_{\rm max}\rangle_d-\langle X_{\rm max}\rangle\right)^2}{ \sigma_X^2}+\frac{\left({\rm Var}(X_{\rm max})_d-{\rm Var}(X_{\rm max})  \right)^2}{ \sigma_{\rm Var}^2}  .
\end{equation}
By minimizing this function with Minuit, the values of $\langle \ln A\rangle$ and Var(ln$A$), together with their uncertainties, can then be obtained. In particular, the values so obtained for the EPOS.LHC-R model adopting the data from \cite{yu19} are those that were used in this paper.

\end{document}